\documentclass[]{aa}  

\usepackage{txfonts}
\usepackage{float}
\usepackage{stfloats}
\usepackage{textcomp}
\usepackage{xcolor}
\usepackage{caption}
\usepackage{graphicx}
\usepackage{hyperref}
\begin{document} 

\newcommand{\periodmasc}{$2.82406\pm {0.00003}$}
\newcommand{\periodmascfinal}{$2.82406\pm 0.000035$}

\newcommand{\epochmasc}{$2458505.817 \pm 0.003$}
\newcommand{\epochmascfinal}{$2458505.817 \pm 0.004$}

\newcommand{\radiusmasc}{$1.53^{+0.07}_{-0.04}$}
\newcommand{\radiusmascfinal}{$1.5\pm 0.1$}

\newcommand{\massmasc}{$3.1 \pm 0.9 $}
\newcommand{\tempmasc}{$2100 \pm 100$}

\newcommand{\durationmasc}{$0.165\pm{0.004}$}
\newcommand{\durationmascfinal}{$0.165\pm 0.006$}

\newcommand{\planettostarratio}{$0.080^{+0.004}_{-0.002}$}
\newcommand{\planettostarratiofinal}{$0.080^{+0.0048}_{-0.0025}$}

\newcommand{\bphotmasc}{$0.4\pm{0.3}$}

\newcommand{\obliquitymasc}{$247.5^{+1.5}_{-1.7}$}
\newcommand{\tempstar}{$7800\pm{200}$}

\newcommand{\semiaxismasc}{$0.047\pm 0.004$}
\newcommand{\semiaxismascfinal}{$0.047 \pm 0.005$}

\newcommand{\stellarmass}{$1.75\pm{0.05}$}
\newcommand{\stellarradius}{$1.92\pm{0.11}$}

\newcommand{\vsinistar}{$46.5 \pm 1$}
\newcommand{\sysvelstar}{$-18.2 \pm 0.2$}
\newcommand{\impactmasc}{$0.34 \pm 0.03$}
\newcommand{\vsini}{$v$sin$i$}
\newcommand{\ms}{m\,s$^{-1}$}
\newcommand{\kms}{km\,s$^{-1}$}
\newcommand{\Msun}{$M_{\odot}$}
\newcommand{\msun}{$M_{\odot}$}
\newcommand{\MSun}{$M_{\odot}$}

   \title{MASCARA-4\,b/bRing-1\,b - A retrograde hot Jupiter around the bright A3V star HD 85628}

   \subtitle{}

   \author{P. Dorval
          \inst{1,2}
          \and
          G.J.J. Talens
          \inst{3}
          \and
          G.P.P.L. Otten
          \inst{4}
          \and
          R. Brahm
          \inst{5,6,7}
          \and
          A. Jordán
          \inst{6,7}
          \and
          L. Vanzi
          \inst{5}
          \and
          A. Zapata
          \inst{5}
          \and
          T. Henry
          \inst{8}
          \and
          L. Paredes
          \inst{9}
          \and
          W.C. Jao
          \inst{9}
          \and
          H. James
          \inst{10}
          \and
          R. Hinojosa
          \inst{10}
          \and
          G.A. Bakos
          \inst{11}
          \and
          Z. Csubry
          \inst{11}
          \and
          W. Bhatti
          \inst{11}
          \and
          V. Suc
          \inst{6}
          \and
          D. Osip
          \inst{12}
          \and
          E.~E. Mamajek
          \inst{13,14}
          \and
          S.~N. Mellon
          \inst{14}
          \and
          A. Wyttenbach
          \inst{1}
          \and
          R. Stuik
          \inst{1,2}
          \and
          M. Kenworthy
          \inst{1}
          \and
          J. Bailey
          \inst{15}
          \and
          M. Ireland
          \inst{16}
          \and
          S. Crawford
          \inst{17,18}
          \and
          B. Lomberg
          \inst{17,19}
          \and
          R. Kuhn
          \inst{17}
          \and
          I. Snellen
          \inst{1}
          }

   \institute{
            Leiden Observatory, Leiden University, Postbus 9513, 2300 RA Leiden, The Netherlands
            \\
              \email{dorval@strw.leidenuniv.nl}
         \and
            NOVA Optical IR Instrumentation Group at ASTRON, P.O. Box 2, 7990 AA Dwingeloo, The
            Netherlands
        \and
            Institut de Recherche sur les Exoplanètes, Département de Physique, Université de Montréal, Montréal, QC H3C 3J7, Canada
        \and
            Aix Marseille Univ, CNRS, CNES, LAM, Marseille, France
        \and
            Center of Astro-Engineering UC, Pontificia Universidad Católica de Chile, Av. Vicuña Mackenna 4860, 7820436 Macul, Santiago, Chile
        \and
            Instituto de Astrofísica, Pontificia Universidad Católica de Chile, Av. Vicuña Mackenna 4860, Macul, Santiago, Chile
        \and
            Millennium Institute for Astrophysics, Chile
        \and 
            RECONS Institute, Chambersburg PA 17201 USA
        \and
            Department of Physics and Astronomy, Georgia State University, Atlanta, GA 30302-4106, USA
        \and
            Cerro Tololo Inter-American Observatory, CTIO/AURA Inc., La Serena, Chile
        \and
            Department of Astrophysical Sciences, Princeton University, 4 Ivy Lane, Princeton, NJ, 08544, USA
        \and
            Las Campanas Observatory, Carnegie Institution of Washington, Colina el Pino, Casilla 601 La Serena, Chile
        \and
            Jet Propulsion Laboratory, California Institute of Technology, 4800 Oak Grove Drive, M/S 321-100, Pasadena, CA, 91109, USA
        \and
            Department of Physics \& Astronomy, University of Rochester, Rochester, NY 14627, USA
        \and
            Department of Physics, University of California at Santa Barbara, Santa Barbara, CA 93106, USA
        \and
            Research School of Astronomy and Astrophysics, Australian National University, Canberra, ACT 2611, Australia
        \and
            South African Astronomical Observatory, Observatory Rd, Observatory Cape Town, 7700 Cape Town, South Africa
        \and
            Space Telescope Science Institute, 3700 San Martin Drive, Baltimore, MD 21218, USA
        \and
            Department of Astronomy, University of Cape Town, Rondebosch, 7700 Cape Town, South Africa
        }

   \date{Submitted 3 April 2019 to A\&A}
% \abstract{}{}{}{}{} 
% 5 {} token are mandatory
\abstract
% {Context heading for MASCARA-4\,b/bRing-1\,b}
{MASCARA and bRing are both instruments with multiple ground-based stations that rely on interline CCDs with wide-field lenses to monitor bright stars in the local sky for variability. MASCARA has already discovered several planets in the Northern sky, which are among the brightest known transiting hot Jupiter systems.
%, ideal for atmospheric characterization.
}
{In this paper, we aim to characterize a transiting planetary candidate in the southern skies found in the combined MASCARA and bRing data sets of HD 85628, an A3V star of V\,=\,8.2 mag at a distance 172 pc, to confirm its planetary nature.}
{The candidate was originally detected in data obtained jointly with the MASCARA and bRing instruments using a BLS search for transit events. Further photometry was taken by the 0.7m CHAT, and radial velocity measurements with FIDEOS on the ESO 1.0m Telescope. High resolution spectra during a transit were taken with CHIRON on the SMARTS 1.5m telescope to target the Doppler shadow of the candidate.}
{We confirm the existence of a hot Jupiter transiting the bright A3V star HD 85628, which we co-designate as MASCARA-4b and bRing-1b. It is in a 2.824 day orbit, with an estimated planet radius of \radiusmasc \space $R_{\rm{Jup}}$ and an estimated planet mass of \massmasc \space $M_{\rm{Jup}}$, putting it well within the planetary regime. The CHAT observations show a partial transit, reducing the probability that the transit was around a faint background star. The CHIRON observations show a clear Doppler shadow, implying that the transiting object is in a retrograde orbit with $|\lambda| = $\obliquitymasc \textdegree. The planet orbits at at a distance of \semiaxismasc \space AU from the star and has a zero-albedo equilibrium temperature of \tempmasc \space K. In addition, we find that HD 85628 has a previously unreported stellar companion star in the Gaia DR2 data demonstrating common proper motion and parallax at 4.3\arcsec separation (projected separation $\sim$740 AU), and with absolute magnitude consistent with being a K/M dwarf.}
{MASCARA-4\,b/bRing-1\,b is the brightest transiting hot Jupiter known to date in a retrograde orbit. It further confirms that planets in near-polar and retrograde orbits are more common around early-type stars. Due to its high apparent brightness and short orbital period, the system is particularly well suited for further atmospheric characterization.}
\maketitle
%
%________________________________________________________________
%\let \cleardoublepage \clearpage

%\newpage
%\pagebreak[4]

\section{Introduction}
\label{Sec: Intro}
%-----------------background----------------------------------

The number of known exoplanets has grown rapidly since the earliest discoveries \citep[e.g.][]{Latham89,Mayor_Queloz1995}, first primarily through radial velocity surveys, and later by successful transit surveys. These initial surveys were performed with small, dedicated ground-based telescopes such as TrES \citep{TrES2007RMxAC..28..129S}, XO \citep{XO2005PASP..117..783M}, HAT \citep{HAT_main2004}, KELT  \citep{KELT2007PASP..119..923P}, and SuperWASP \citep{SuperWASP_main2006}. The space-based transit mission CoRoT \citep{CoRoT2008}, and in particular Kepler \citep{Kepler_main}, have increased the exoplanet tally to several thousand. These surveys indicate that, on average, there is at least one planet orbiting every late-type main sequence star \citep{Batalha2014}.

Hot Jupiters $-$ gas giant planets that closely orbit their host star $-$ are relatively rare, but over-represented in both radial velocity and transit surveys because they are the easiest to find 
\citep{Wright12,Dawson18}. They are unlikely to be formed in-situ due to the close proximity to their host star, and must have migrated from their formation location due to either interactions with the circumstellar disk, or with other bodies in the system. Measurements of the projected angle between the stellar spin axis and the planet orbital plane are accessible through the Rossiter-McLaughlin effect (also called the Doppler shadow) and may point to a mix of migration scenarios. Their large sizes, high effective temperatures, and high transit probability $-$ with transits and eclipses occurring frequently if observed near edge-on $-$ make them ideal targets for atmospheric follow-up and characterization. Their high temperatures imply that their atmospheres could be well mixed, providing means to compare their chemical composition to that expected from different formation locations in the protoplanetary disk. 
Hot Jupiters are also prone to atmospheric evaporation and escape \citep{escape12003Natur.422..143V,evaporate2003ApJ...598L.121L}, processes that were likely important in the early solar system, including Earth.
In any case, hot Jupiters will always remain the easiest exoplanet targets to characterize, meaning that we will get the most detailed observational information, challenging our modeling and understanding to the extreme $-$ now and in the future. 

A key parameter for exoplanet atmospheric characterization is the apparent brightness of the system. Kepler, and to lesser extent its successor K2, only targeted very faint stars, while most of the brighter transiting hot Jupiter systems have been found with dedicated ground-based transit surveys. The two best studied hot Jupiters, HD 209458\,b \citep{HD209458_2000ApJ...529L..45C} and HD 189733\,b \citep{HD189733_2005A&A...444L..15B} were discovered via radial velocity searches and found to transit later. Systems brighter than $m_V \sim 8$ saturate most ground-based searches, although recently both SuperWASP and the KELT survey have discovered very bright systems by altering their survey strategy.

The Multi-Site All-Sky CAmeRA, MASCARA \citep{MASCARA_instrument,MASCARA_instrument2} is specifically designed to find the brightest transiting hot Jupiters in the sky. It has so far found MASCARA-1 b \citep{MASCARA_science_1}, MASCARA 2-b \citep{MASCARA_science_2}, and MASCARA 3-b \citep{MASCARA_Science3}. MASCARA is combined with the bRing network \citep{bRing_instrument}, which is based on similar technology as MASCARA. 
%Its main science case was to study the Hill-sphere transit of $\beta$ Pictoris\,b \citep{MellonAAS2019, KalasAAS2019}, but can also be used for exoplanet transit searches and for variable star characterization, as with the young hybrid debris disk star HD 156623 \citep{Mellon2019}.
bRing's main science goal was to study the Hill-sphere transit of $\beta$ Pictoris\,b \citep{MellonAAS2019, KalasAAS2019}; bRing has also been used for exoplanet transit searches (this work) and variable star characterization \citep[e.g.,][]{Mellon2019}.

In 2018, NASA launched the Transiting Exoplanet Survey Satellite (TESS) \citep{TESS_main}, aimed to find a wide range of transiting planets around bright stars \citep{Tess_ExoplanetPredictions}, including those targeted by MASCARA and bRing.   
In this paper we present the discovery of MASCARA-4\,b/bRing-1\,b (further referred to as MASCARA-4\,b), soon to be observed by TESS. MASCARA-4\,b orbits the bright ($m_V=8.2$) A3V star HD 85628, which also possesses a previously unreported dim binary companion. In Section \ref{Sec: Discovery_Obs} we describe the discovery observations performed by the MASCARA and bRing network. Section \ref{Sec: Followup} describes the follow-up observations performed by CHAT, FIDEOS, and CHIRON. Section \ref{Sec: Parameters_Analysis} presents the full analysis of the system parameters of MASCARA-4\,b, which are discussed in Section \ref{Sec:Discussion}.

%__________________________________________________________________

\section{MASCARA, bRing, and Discovery Observations}
\label{Sec: Discovery_Obs}

The primary objective of MASCARA \citep[]{MASCARA_instrument2} is to find transiting planetary systems around bright ($4 < m_V < 8$) stars. It consists of two stations, a northern station at the Observatorio del Roque de los Muchachos, La Palma, Canary Islands in Spain, which has been observing since 2015, and a southern station at the European Southern Observatory (ESO) at La Silla in Chile, which has been observing since mid-2017. Each station is equipped with five interline CCD cameras with wide-field lenses that allow each station to observe the local sky down to airmass $\sim$2 \citep{MASCARA_instrument2}. As each camera is observing a fixed direction, stars are moving across the same track on the CCDs during each night. Each image is taken with an exposure time of 6.4 seconds, which is short enough such that stars travel over less than a pixel during one exposure. Photometry from every 50 exposures are binned together after reduction and calibration \citep{MASCARA_calibration}. As the detectors are interline CCDs, the readout of each image is performed during the next exposure. Hence, no observing time is lost between exposures. This requires large data storage capabilities, as each station generates $\sim$15 TB of data each month. For economic and practical reasons, the basic data reduction steps are performed on site, with the raw data being overwritten after several weeks. A permanent record of the sky is kept in the form of stacked images, which can be used for future searches of transients and transits. 

The MASCARA stations are paired with bRing, two photometric instruments that were constructed to observe $\beta$ Pictoris during the expected Hill sphere transit of $\beta$ Pictoris b, to search for a possible giant exoring system \citep{Kenworthy2017NatAs...1E..99K,bRing_instrument, MellonAAS2019, KalasAAS2019}. 
bRing consists of two stations in the southern hemisphere, one at the South African Astronomical Observatory (SAAO) at Sutherland in South Africa, and the other at the Siding Springs Observatory (SSO) in Coonabarabran, Australia. Each bRing station was designed along the same basic principles as MASCARA, and is equipped with two wide-field lenses with interline CCD cameras of the same model as the southern MASCARA station. With two cameras, each bRing can only see between a declination range of -30\textdegree\, to -90\textdegree, but observe from horizon to horizon in the east-west direction down to airmass 10. bRing operates at consecutive exposures of 6.4 and 2.6 seconds to prevent saturation on the bright star $\beta$ Pictoris b, with the photometry from 25 long and 25 short exposures being binned together after reduction and calibration. This allows for the combination of data between the two bRing stations and the southern MASCARA station.

\begin{figure}[ht]
    \centering
    \includegraphics[scale=0.25]{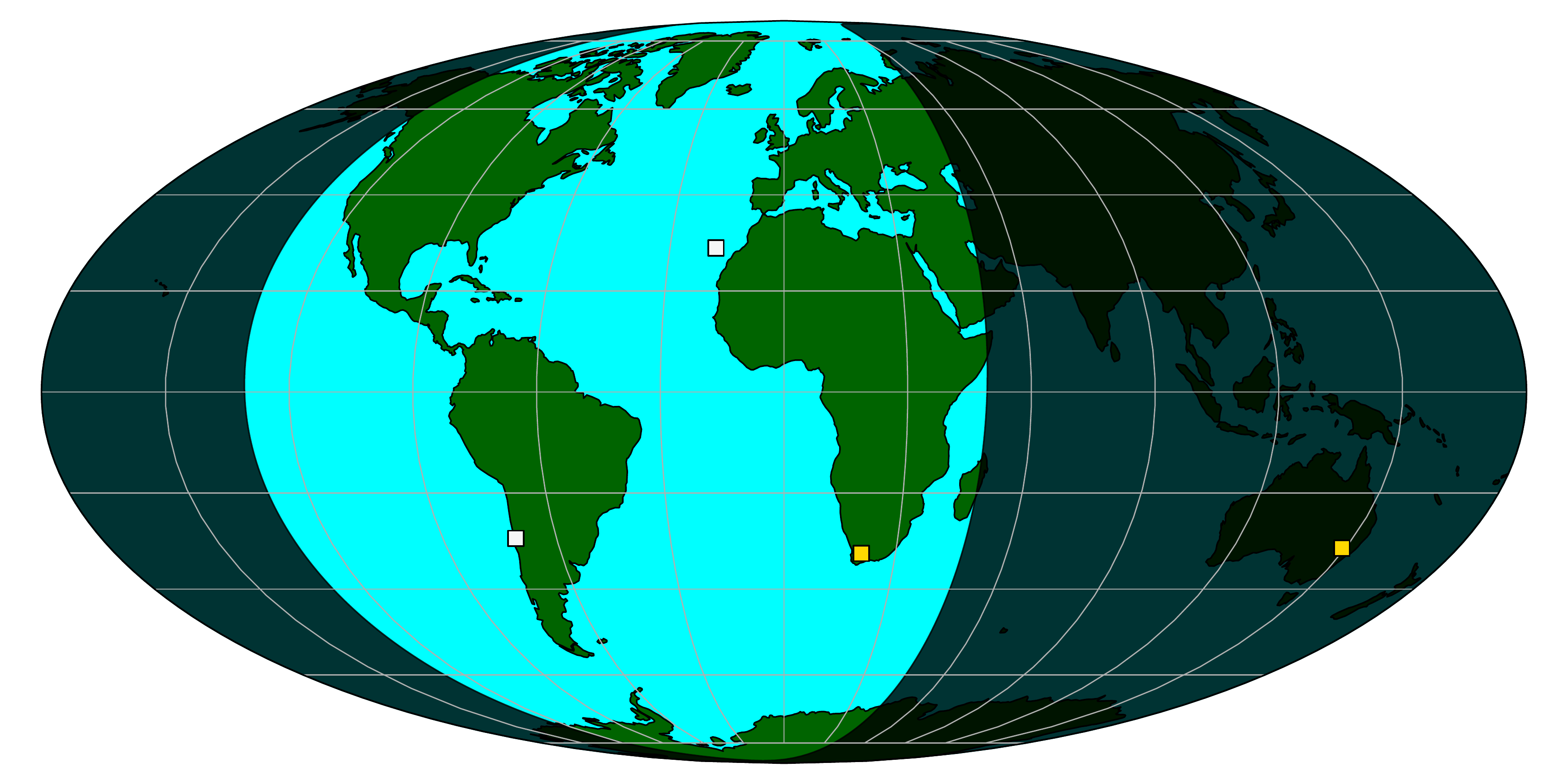}
    \caption{Global view of the MASCARA/bRing network. The two MASCARA instruments (white squares) are located in La Silla, Chile, and La Palma, the Canary Islands, Spain. The two bRing instruments (yellow squares) are located in Sutherland, South Africa, and Coonabarabran, Australia. At any given time, at least one instrument in the southern hemisphere is observing, provided weather permits.}
    \label{fig:MASCARA_Network}
\end{figure}

With the presence of three observatories in the southern hemisphere, the MASCARA/bRing network allows for continuous observations of stars within the declination range $-30$\textdegree $>\delta >-90$\textdegree \space(local weather permitting). As shown in Fig. \ref{fig:MASCARA_Network}, the observatories are placed to obtain as much continuous observation time as possible. Throughout most of the year, weather permitting, at least one observatory is exposing at any given time. At worst, during the summer there is a gap of thirty to sixty minutes between stations. This continuity of observations is beneficial to transit searches, in particular for planets with orbital periods close to multiples of one day. For a detailed description of the calibration procedure of MASCARA and bRing, we refer the reader to \citet{MASCARA_calibration}. 

\begin{figure*}[h!]
 \centering
 \includegraphics[width=0.97\textwidth]{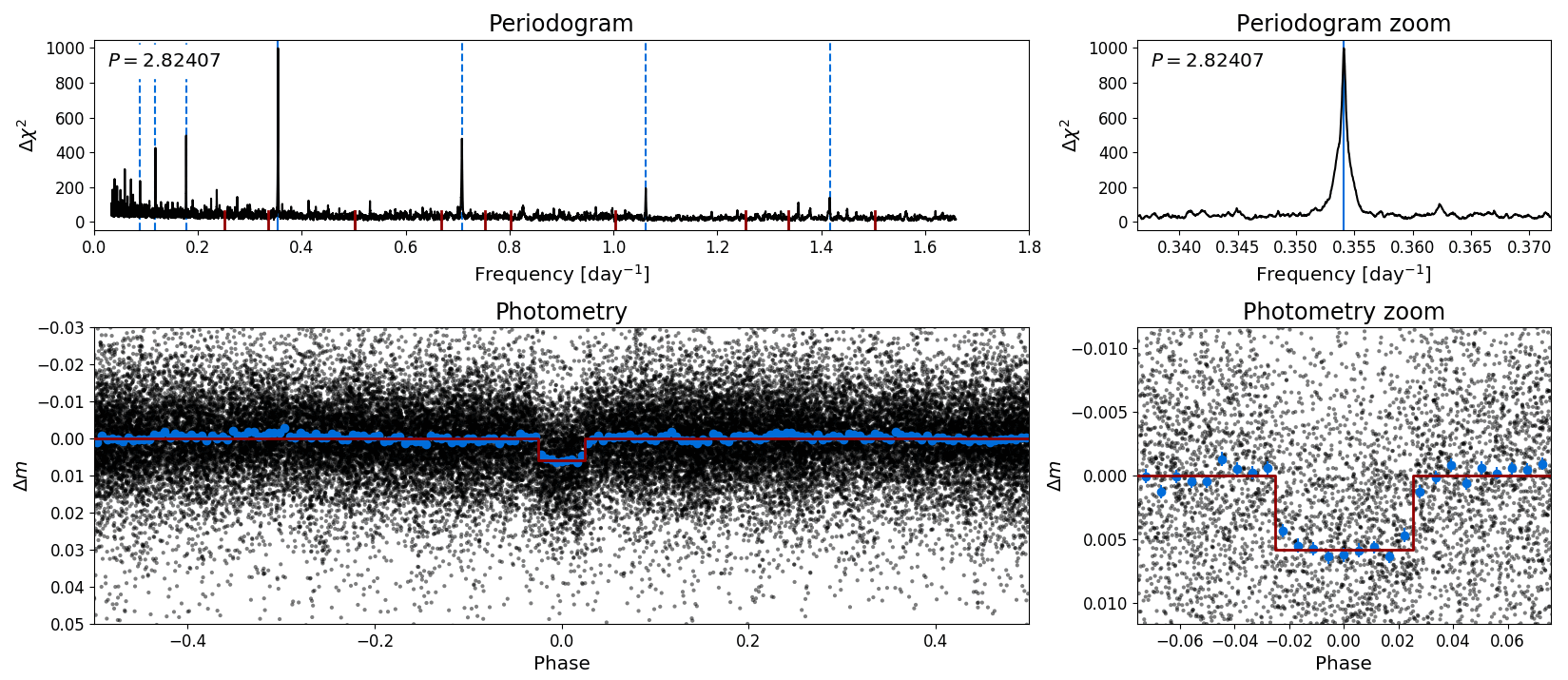}
 \caption{Discovery data of MASCARA-4\,b/bRing-1\,b. \textit{Top left}: BLS periodogram of the combined MASCARA and bRing photometry obtained between mid-2017 and end-2018. \textit{Top right}: The same zoomed in on the peak in the periodogram, which is located at 0.354 days$^{-1}$. \textit{Bottom left}: The calibrated MASCARA and bRing data, phase folded to a period of 2.82407 days. The blue dots are binned such that there are 9 data points in transit, which comes out to phase steps of $\sim$ 0.005. \textit{Bottom right}: The same zoomed in on the transit event.}
 \label{fig:MASCARA_Discovery_Transit}
\end{figure*}

A Box Least-Square analysis \citep[BLS]{BLSKovacs2002A&A...391..369K} on the combined lightcurve for HD 85628 reveals a strong signal at a period of \periodmasc \space days. The BLS periodogram, and the phase-folded lightcurve of the MASCARA and bRing data are shown in 
Fig. \ref{fig:MASCARA_Discovery_Transit}. The combined light curve of HD 85628 consists of 52296, 320 second binned data points, totalling 4500 hours of data. 

\section{Follow-up Observations}
\label{Sec: Followup}
After the initial detection of the planet-like signal with MASCARA and bRing, additional follow-up observations were taken to confirm the transit and planetary nature, and constrain the mass of the planet. Photometric observations were taken with the Chilean-Hungarian Automated Telescope (CHAT) to reduce the possibility that the transit signal originates from a faint background star. Radial velocity measurements were taken with FIDEOS on the ESO 1m Telescope to constrain the mass of the transiting object to the planetary regime. High-resolution spectra were taken during transit with the CHIRON instrument on the SMARTS telescope to detect the Doppler shadow of the transiting object. This provides conclusive evidence that the object is indeed transiting the bright star, and in combination with the radial velocity data, is of planetary nature. In addition, it provides the projected spin-orbital angle of the system. Table \ref{Tab: followup Obs} details all photometric and spectroscopic observations used to discover and confirm MASCARA-4\,b.  

\begin{table*}[h]
\centering
\caption{Observations used in the discovery and confirmation of MASCARA-4\,b/bRing-1\,b. Listed dates are in the format dd-mm-yyyy.}
\begin{tabular}{lllrrl}
Instrument & Observatory & Date & $N_{\rm{obs}}$ & $t_{\rm{exp}}$ [s] & Filter/Spectral Range \\ \hline \hline
MASCARA-S & La Silla & 01-10-2017 - 31-12-2018 & 15879 & 320 &None \\ 
bRing-SA & SAAO & 01-07-2017 - 31-12-2018 & 19672 & 320 &None \\
bRing-AU & SSO & 01-10-2017 - 31-12-2018    & 16745 & 320 &None \\ 
FIDEOS & La Silla & 24-05-2018 - 27-05-2018  & 14 & 900 & 400-700 nm \\
FIDEOS & La Silla & 06-11-2018 - 21-11-2018 & 13 & 900 & 400-700 nm \\
CHAT & Las Campanas & 31-10-2018 & 271 & 4 & 700-810 nm \\
CHIRON & CTIO & 25-01-2019 & 66 & 240 & 460-875 nm \\
\end{tabular}
\label{Tab: followup Obs}
\end{table*}

\subsection{Photometric measurements with CHAT}
\begin{figure}[h]
    \centering
    \includegraphics[scale=0.5]{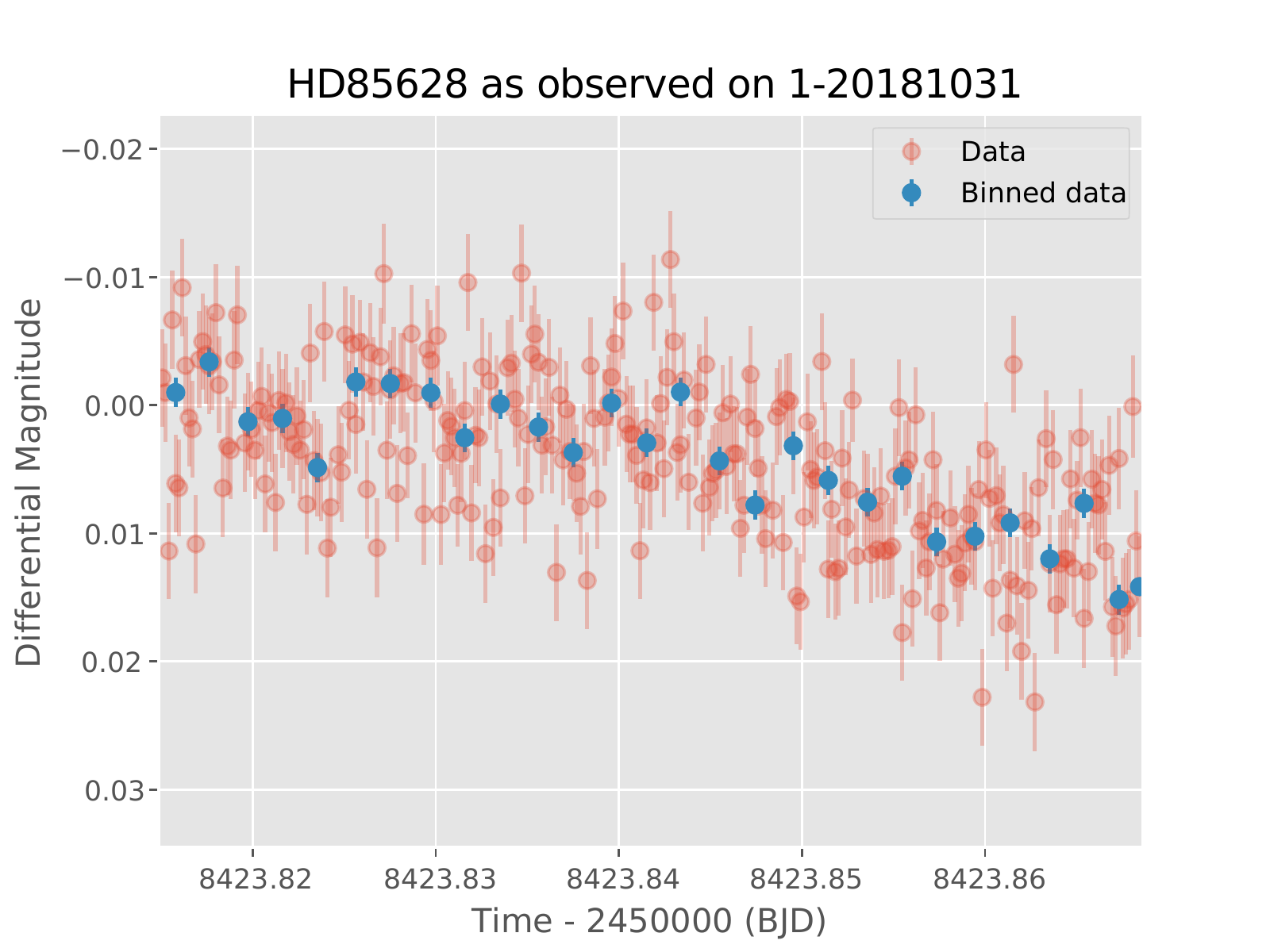}
    \caption{CHAT follow-up observations of MASCARA-4\,b. An ingress is clearly seen which is consistent with the ephemeris and transit shape determined from the MASCARA/bRing data.}
    \label{fig:CHAT_LC}
\end{figure}

MASCARA-4 was photometrically monitored with the 0.7m Chilean-Hungarian Automated Telescope (CHAT), located at Las Campanas Observatory, on 31 October 2018, during a transit event. Observations were performed with the $i'$ filter and a slight defocus was applied. The exposure time of each image was set to 4 seconds which produced a typical peak flux of 10,000 ADUs per exposure. CHAT data was processed with a dedicated pipeline \citep[see][]{hartman:2019,jordan:2018,espinoza:2018} that performs the classical data reduction, the aperture photometry, and the generation of the light curve by selecting the more suitable comparison stars and photometric aperture (12 pixels or 7" in this case) for the differential photometry. Fig. \ref{fig:CHAT_LC} shows the CHAT light curve, which confirmed the presence of a transit like feature on MASCARA-4 by showing a definite ingress of the possible planetary companion. The timing and depth of this partial transit is consistent with the ephemeris determined from the MASCARA/bRing data. It reduces the probability of the transit signal originating from a faint background star by orders of magnitude. Since the CHAT lightcurve only partially covers the transit, it is not used to constrain the orbital period and other transit parameters. 

\subsection{Radial velocity measurements with FIDEOS}

We obtained twenty-seven spectra of MASCARA-4 with the FIber Dual Echelle Optical Spectrograph \citep[FIDEOS,][]{vanzi:2018} mounted on the ESO 1.0 m telescope at La Silla Observatory. FIDEOS has a spectral resolution of 42,000, and covers the wavelength range from 400 nm to 700 nm.
It is stabilized in temperature at the 0.1 K level, and it uses a secondary fibre illuminated with a ThAr lamp for tracing the instrumental radial velocity drift during a scientific exposure. Fourteen spectra were acquired during UT 24-27 May 2018, while another set of thirteen spectra were obtained between UT 6-21 November 2018. The adopted exposure time for each measurement was 900 seconds, which generated a typical SNR at 5150 \AA \space of 105. FIDEOS data were processed with a dedicated automatic pipeline built using the routines from the CERES package \citep{brahm:2017}. This pipeline performs the frame reductions, the optimal extraction of each spectra, the wavelength calibration, and the instrumental drift correction. Dedicated IDL procedures were used to derive the radial velocity variations of the target. 

To use a cross-correlation template, a detailed synthetic spectrum was computed using the IDL interface SYNPLOT (I. Hubeny, private communication) to the spectrum synthesis program SYNSPEC\footnote{http://nova.astro.umd.edu/Tlusty2002/pdf/syn43guide.pdf}, utilizing Kurucz model atmospheres\footnote{http://kurucz.harvard.edu/grids.html}.

\subsection{Doppler-shadow measurements with CHIRON}

High-resolution spectroscopic data were taken via the CTIO high resolution spectrometer (CHIRON) on the SMARTS 1.5 m telescope in Chile. MASCARA-4 was observed on UT 25 January 2019 during one of the predicted transits, in order to measure the Rossiter-McLaughlin effect which would be an unambiguous signature of the planetary nature of MASCARA-4\,b. We obtained a series of 60 spectra, 50 being in-transit and 10 out-of-transit. The spectra were reduced according to the CHIRON pipeline. The start of the transit was missed, and observations continued for around thirty minutes after end-of-transit.

\section{System and Stellar Parameters \label{Sec: Parameters_Analysis}}
\subsection{Photometric transit fitting}

The photometric modelling of the transit is performed in a similar way as for MASCARA-1 b \citep{MASCARA_science_1} and MASCARA-2 b \citep{MASCARA_science_2}. We fit a \citet{MandelAlgol2002ApJ...580L.171M} model to the MASCARA data using a Markov-chain Monte Carlo (MCMC) approach, using the Python codes \textsc{batman} \citep{BATMAN2015PASP..127.1161K} and \textsc{emcee} \citep{EMCEE2013PASP..125..306F}. We assumed a circular orbit ($e=0$), and optimized for the fit parameters: the transit epoch $T_P$, the orbital period $P$, the transit duration $T_{14}$, the planet-to-star radius ratio $p$, and the impact parameter $b$. We use parameters obtained from the discovery BLS algorithm as initial parameters. We fit the uncorrected MASCARA data, including polynomials used to correct instrumental effects \citep{MASCARA_calibration}. For this fit, a linear limb-darkening coefficient of 0.6 was used. The best-fit parameter values come from the median of the output MCMC chains, and the 1$\sigma$ uncertainties from the 16th and 84th percentiles of the distributions.

\begin{figure}
 \centering
  \includegraphics[scale=0.4]{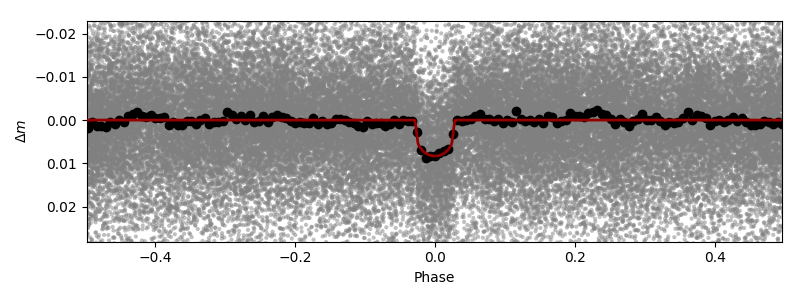}
 \caption{The photometric lightcurve of MASCARA-4\,b, using the best-fit parameters listed in Table \ref{Tab: Photometric_Fit}. The grey points indicate individual data, and the black dots are binned the same as in Fig. \ref{fig:MASCARA_Discovery_Transit}. The red line shows the photometric model.}
 \label{fig:MASCARA_Photometric Analysis}
\end{figure}

Table \ref{Tab: Photometric_Fit} lists the parameters derived from this model, as well as the reduced chi-square of the data. The reduced chi-square of 1.43 of this fit indicates that the errors are likely underestimated by about 20\%. Fig. \ref{fig:MASCARA_Photometric Analysis} shows the phase-folded photometric lightcurve of MASCARA-4\,b with the best-fit transit model.

\begin{table*}[h!]
\centering
\caption{Parameters and best-fit values derived from the MASCARA and bRing joint photometric data. It is important to note that the reduced chi-square indicates the errors are likely underestimated by $~20\%$. }
\begin{tabular}{rccr}
Parameter & Symbol & Units & MASCARA\\ \hline \hline
Reduced chi-square &  $ \chi _\nu^2$ & - & 1.43 \\
Epoch& $T_P$ & BJD & \epochmasc \\
Period& $P$ & days &  \periodmasc \\
Duration& $T_{14}$ & days & \durationmasc \\
Planet-to-star ratio & $R_P/R_{\ast}$ & - & \planettostarratio\\
Impact parameter & $b$ & - & \bphotmasc \\
Eccentricity & $e$ & - &  0 (fixed) \\
\hline
\end{tabular}
\label{Tab: Photometric_Fit}
\end{table*}

\subsection{Radial velocity analysis and planet mass}

\begin{figure}
    \centering
    \includegraphics[width=0.48\textwidth]{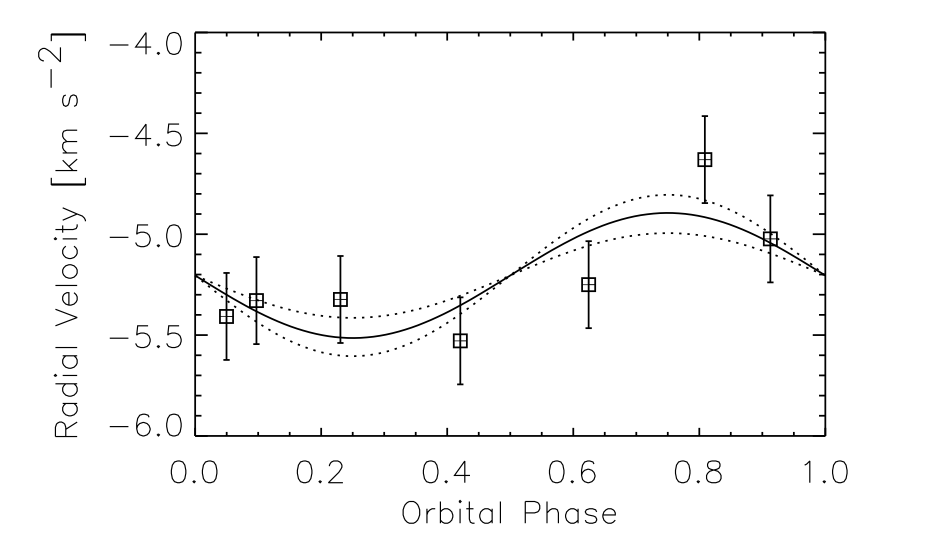}
    \caption{Radial velocity data of MASCARA-4 taken with the FIDEOS spectrograph on the ESO 1m telescope. A marginal sinusoidal variation is detected at 3$\sigma$ with an amplitude of $310 \pm 90$ \ms. Assuming a stellar mass of \stellarmass \space M$_\odot$, this corresponds to a planet mass of \massmasc M$_{\rm{Jup}}$, well within the planetary regime.}
    \label{RVpoints}
\end{figure}

For the RV analysis we used the twenty-seven FIDEOS spectra. 
Since the projected equatorial rotation velocity of the star is very high, rotational broadening is the dominant component in the line-profiles, resulting in substantial uncertainties on the RV measurements. 
The individual RV points were phase folded using the best-fit values from photometry, and individual data points taken in sequence on the same night were binned, resulting in seven RV measurements with uncertainties of $\sim$200 m s$ ^{-1} $ (see Fig. \ref{RVpoints}).

A circular orbital solution, with RV amplitude and system velocity as free parameters, was fit to the data. This results formally in a detection of a sinusoidal variation with an amplitude of $310 \pm 90$ \ms. Assuming a stellar mass of \stellarmass \space M$_\odot$ (see \S4.4), this corresponds to a planet mass of \massmasc \space M$_{\rm{Jup}}$, well within the planetary regime.

\subsection{Doppler-shadow analysis}
\label{Subsec:DopplerAnalysis}
In order to successfully validate the planetary nature of MASCARA-4\,b, a spectroscopic time series during a transit was taken with CHIRON. This observation allows us to confirm that a planet-sized dark object transits the bright, fast spinning star. In addition, the projected spin-orbit angle of the system is determined. For our analysis, Cross Correlation Functions (CCF) were constructed from the reduced CHIRON spectra using the same Kurucz spectral template as used for the radial velocity analysis. In our analysis, we used all 41 orders blueward of 6950 \AA, except orders 7, 8, 28, 29, 34, 37, 38 to avoid the Balmer series and telluric contaminations. The CCF for each observation was constructed by summing the CCF of each individual order, which were subsequently all scaled to the same level. 

\begin{figure*}
    \centering
    \includegraphics[width=0.97\textwidth]{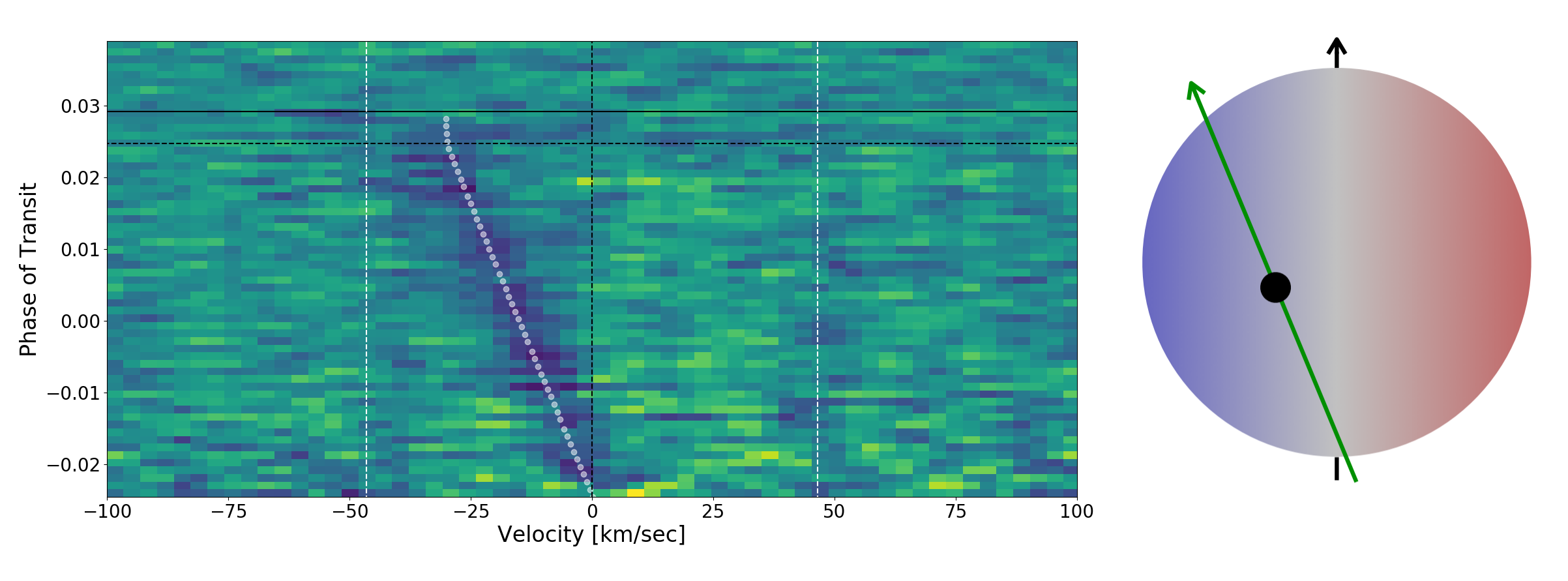}
    \caption{\textit{Left:} Doppler-shadow measurements on MASCARA-4\,b during transit as derived from observations taken with the CHIRON spectrograph on the SMARTS 1.5m telescope. These consist of sixty mean stellar line profiles from which the average out-of-transit profile is subtracted, showing the movement of the dark planet in front of the fast-rotating star. The black horizontal lines denote t$_3$ (dashed, the time when the orbiting object starts to egress) and t$_4$ (solid, when the orbiting object is no longer transiting) of the transit. The dashed vertical black line represents the systemic velocity of the star, which is set to 0\,\kms. The two vertical white dashed lines indicate the extent of the velocity broadened stellar line profile with a \vsini\space of 46.5\,\kms. The white dotted line represents the best-fit model, corresponding to an impact parameter of \impactmasc, and $|\lambda|$ of \obliquitymasc. \textit{Right:} Visualization of the orientation of the planet orbit with respect to the fast spinning star. }
    \label{fig:Doppler_Shadow}
\end{figure*}

In order to extract the Rossiter-McLaughlin information from the data, we followed the ``Reloaded'' method \citep{Cegla2016b,Bourrier2016b,Bourrier2018a,Wyttenbach2017}. We first determined the average stellar CCF by taking the mean of the out-of-transit CCFs, corresponding to the last ten observations. Then, we normalised each CCF by the BLS photometric fit, and subsequently computed the difference between the out-of-transit CCF and each observation. The final residual CCFs are obtained by removing the constant offset. The Doppler shadow of the planetary sized object, coined as MASCARA-4/,b is clearly seen in the data (Fig. \ref{fig:Doppler_Shadow}). Note that the ingress of the transit was just missed. A Gaussian model was fitted to the residual CCFs to determine the radial velocity of the Doppler shadow feature at each epoch during the transit, while the stellar systemic velocity of $v_{\rm{sys}} = $ \sysvelstar\ km~s$^{-1}$ was removed from each measurement. Finally, 34 out of the 50 observations in transit had a sufficient velocity precision in order to be used to constrain the Rossiter-McLaughlin effect and to obtain $\lambda$, the projected spin-orbit angle. To model the residual CCF velocities, we computed the brightness-weighted average rotational velocity behind the planet during each exposure as a function of the transit parameter, the spin-orbit angle, and the stellar velocity field which is given, when assuming a rigid body rotation, by the \vsini \space \citep{Cegla2016b}. As no high-precision photometric light curve was available, the impact parameter was poorly determined. Thus, instead of fitting a model for $\lambda$ and \vsini, we decided to fix the value of \vsini \space to its spectroscopic measurement of \vsinistar\ \kms\, and to fit for the impact parameter and $\lambda$. Practically, we allowed the measured \vsini\space to vary within its 3$\sigma$ uncertainty. As the impact parameter and $\lambda$ are both correlated with the \vsini, the allowed range in \vsini\, gives us the 1$\sigma$ uncertainty on the impact parameter and $\lambda$, while their best values are fitted with a $\chi^2$ minimisation. The resulting fit is shown as a white dotted line in Fig. \ref{fig:Doppler_Shadow}. We find an impact parameter of \impactmasc\ and a spin-orbit angle $|\lambda|$ of \obliquitymasc \textdegree. Such an obliquity is set as retrograde, but is at the edge of being classified as polar (We use the standards present in \cite{Addison2013ApJ...774L...9A}, with a retrograde orbit being classified as having an obliquity of $112.5$\textdegree$\leq\  |\lambda|\leq247.5$ \textdegree, and a polar orbit being classified as having an obliquity of $67.5$\textdegree$<  |\lambda|<112.5$ \textdegree). This makes MASCARA-4\,b the brightest known system with a hot Jupiter in a retrograde orbit. A visualization of the projected orientation of the orbit is shown in the right panel of Fig. \ref{fig:Doppler_Shadow}.

\subsection{Stellar Parameters}

HD 85628 is a relatively unstudied V=8.19 star in the Carina constellation \citep{ESA97}, with its only previous reference in SIMBAD being 
from spectral classification in the Michigan Spectral Survey \citep[A3V;][]{Houk75}. 
The star's Gaia DR2 trigonometric parallax ($\varpi$ = 5.8297\,$\pm$\,0.0318 mas) implies a distance of $d$ = 171.54\,$\pm$\,0.94 pc. 
HD 85628's astrometric and photometric observables are summarized in Table \ref{Tab:stars}.
The stars optical and near-IR colors are consistent with a A7V \citep{Pickles10}, and confirmed via comparison of the
observed colors (e.g. $Bp-Rp$ = 0.2548, $B-V$ = 0.20, $V-Ks$ = 0.45) to those of main sequence dwarfs in the table 
by \citep{Pecaut13}\footnote{See updated table at: http://www.pas.rochester.edu/$\sim$emamajek/ EEM$\_$dwarf$\_$UBVIJHK$\_$colors$\_$Teff.txt.} 
From the {\it Stilism} 3D reddening maps of the solar vicinity by \citet{Capitanio17}, we estimate that a star at 
HD 85628's position would have reddening E($B-V$) = 0.020\,$\pm$\,0.018 mag, which for the standard reddening law \cite{Fiorucci03} translates
to an extinction of A$_V$ = 0.063\,$\pm$\,0.056 mag.
We discount the large extinction and reddening values quoted in the Gaia DR2 catalog \citep{GaiaDR2} ($A_G$ = 0.5012$^{-0.2262}_{+0.1438}$ mag,
  E($B_P-R_P$) = 0.2580$^{+0.0663}_{-0.1097}$ mag), which would unphysically require the star to really be a $\sim$A0V star\footnote{
  Unusually high reddenings and extinctions from Apsis-Priam appear to be a feature of the Gaia DR2 catalog. The
  well-characterized stars with SIMBAD entires with DR2 parallaxes that place them within 10\,pc have a median extinction
  of $A_G$ $\simeq$ 0.24 mag (mean 0.34, standard deviation 0.39 mag) in the Gaia DR2. These stars should all comfortably lie within the Local 
  Bubble with near-zero extinction \citep[][]{Reis11}.}.

We further constrain the stellar parameters for HD 85628 using the Virtual Observatory SED Analyzer (VOSA)\footnote{VOSA 6.0: http://svo2.cab.inta-csic.es/theory/vosa/} and fitting the H$\alpha$ profile of the star's CHIRON spectrum. We fit HD 85628's 
spectral energy distribution using published photometry from Tycho \citep[$B_T V_T$;][]{ESA97}, 
APASS \citep[$BV$;][]{Henden16}, Gaia DR2 \citep[$B_p R_p G$;][]{GaiaDR2}, DENIS \citep[$JK$;][]{DENIS}, 2MASS \citep[$JHK_s$;][]{Skrutskie06}, and WISE \citep[$W1 W2 W3 W4$;][]{Wright10}, and fit Kurucz ATLAS9 stellar atmosphere models \citep{CastelliKurucz}. Given that the star is clearly main sequence and relatively young ($<$1 Gyr), we constrain the surface gravity to be within $\pm$0.5 dex of log$g$ = 4.0 and metallicity within $\pm$0.5 dex of solar, and include the extinction and 1$\sigma$ uncertainty presented previously as a constraint, but allow the effective temperature to float. We find the best fit parameters to be for a Kurucz model with T$_{eff}$, = 7844$^{+57}_{-285}$ K,
log$g$ = 4.0 and solar metallicity. We also fit the shape of the H$\alpha$ line from the average high-resolution CHIRON spectra with a grid of Kurucz spectra (as used above). While we fail to meaningfully constrain its metallicity and surface gravity in this way, the spectral line shape is consistent with a temperature of 7700\,$\pm$\,300 K. Based on these independent analyses, we
adopt T$_{\rm eff}$ $\simeq$ 7800\,$\pm$\,200\,K. We note that this is somewhat cooler than expected given
the A3V classification by \citet{Houk75}, since typical A3V stars have T$_{\rm eff}$ $\simeq$ 8550\,K \citep[][]{Pecaut13}, however we can not reconcile such a hot temperature given the available data.
The best fit luminosity to SED fit using VOSA, and adopting the distance based on the Gaia DR2
trigonometric parallax, is $L$ = 12.23\,$\pm$\,0.0655 $L_{\odot}$ or log($L/L_{\odot}$) $\simeq$ 1.087\,$\pm$\,0.023 dex. For the adopted 
effective temperature, this implies that HD 85628 has a radius of 1.92\,$\pm$\,0.11 R$_{\odot}$
\footnote{Adopting IAU 2015 nominal solar values L$_{\odot}$ 
= 3.828$\times$10$^{26}$ W and R$_{\odot}$ = 695700 km.}.\\

\begin{figure*}[htbp]
    \centering
    \includegraphics[scale=0.6]{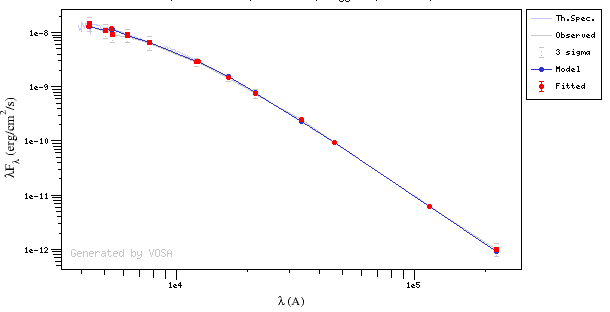}
    \caption{Spectral energy distribution for HD 85628 generated using VOSA. Photometry from Tycho,  Gaia DR2, APASS, DENIS, 2MASS, and WISE are plotted (references in \S4.4). Overlain is a synthetic stellar spectrum from \citet{CastelliKurucz} for T$_{\rm eff}$ = 7750\,K with log$g$ = 3.5, solar metallicity, $A_V$ = 0.07, and corresponding to bolometric luminosity 12.23\,$\pm$\,0.655 L$_{\odot}$ (adopting the distance implied by the inverse of the Gaia DR2 trigonometric parallax).  
    \label{fig:SED}}
\end{figure*}

The duration of the transit provides an estimate of the average density of the host star, assuming a circular orbit and providing that the impact parameter is well known. The latter is very well constrained ($b$\,=\,\impactmasc) by the Doppler shadow measurements, with the T$_{14}$ transit duration measured to be \durationmasc \space days. This implies that if the impact parameter would have been 0, the transit duration would have been $T_{\rm{center}} / \sqrt{1 - b^2}$ = 0.175$\pm$0.005 days, corresponding to the planet travelling over a distance of $2( R_* + R_{\rm{p}}) = 2.16 \pm 0.006$R$_*$. By feeding this into Kepler’s third law, we obtain a mean stellar density of 0.41\,$\pm$\,0.04 g cm$^{-3}$ (0.29\,$\pm\,0.03 \, \rho_{\odot}$).

Interpolating the solar metallicity evolutionary tracks and isochrones of \citet{Bertelli09} using our adopted effective temperature and luminosity, one predicts a mass of 1.75\,$M_{\odot}$, age of 0.82 Gyr,  and surface gravity log$g$ = 4.11. We use the online PARAM 1.1 interface\footnote{http://stev.oapd.inaf.it/cgi-bin/param$\_$1.0} to provide Bayesian estimates of
stellar parameters based on the HR diagram position for HD
85628. Using the evolutionary tracks of \citet{Girardi00}, and
adopting $A_V$ = 0.07 and metallicity within $\pm$0.1 dex of solar,
PARAM predicts an isochronal age of 0.7\,$\pm$\,0.2 Gyr, mass
1.753$\pm$0.044 $M_{\odot}$, log$g$ = 4.09$\pm$0.05.
Given the empirical mass-luminosity relationship for well-measured main sequence
stars by \citet{Eker15}, our adopted $L$ for HD 85628 would predict a mass of 1.78\,$\pm$\,0.11 M$_{\odot}$. Based on these estimates, we adopt a mass of 1.75\,$\pm$\,0.05 $M_{\odot}$, log$g$ $\simeq$ 4.1, and age 0.8 Gyr. The predicted density (0.252\,$\pm\,0.045 \rho_{\odot}$) compares well to that predicted by the transit and Kepler's third law (0.29\,$\pm\,0.03 \rho_{\odot}$)\\

\subsubsection{Stellar Companion HD 85628 B}

The Gaia DR2 catalog shows that the star HD 85628 (Gaia DR2 5245968236116294016) has a companion, which we 
designate HD 85628 B (Gaia DR2 5245968236111767424), previously unreported in the Washington Double Star Catalog 
\citep[]{Mason01}. The companion's $G$-band magnitude (14.0490\,$\pm$\,0.0047) is 5.875 mag fainter 
than that of HD 85628, and it lies 4336.21\,$\pm$\,0.06 mas away at position angle 224$^{\circ}$.938 (ICRS, epoch 2015.5), 
corresponding to a projected separation of $\sim$736 AU. The pair clearly demonstrate common proper motion and parallax 
as seen in Table \ref{Tab:stars}. The proper motion of B with respect to A is $\Delta$$\mu_{\alpha}$, $\Delta$$\mu_{\delta}$
= -0.20\,$\pm$\,0.09, +2.15\,$\pm$\,0.08 mas yr$^{-1}$, which at the weighted mean distance (170.0\,$\pm$\,0.7 pc) 
corresponds to a tangential (projected orbital motion) velocity of B with respect to A of 1.74\,$\pm$\,0.09 \kms. 

HD 85628 B's Gaia DR2 $B_p$ and $R_p$ photometry are likely corrupted given the
large value of E(BR/RP) [2.325], which far exceeds the threshold for reliable photometry
for stars of similar $B_p$-$R_p$ color determined by \citet{Evans18}. 
Besides Gaia DR1 and DR2, the only pre-Gaia catalog that has an entry for HD 85628 B
is USNO-B1.0 \citep{Monet03} (USNO B1.0-0238-0154014).
We have limited color data and no spectral data for HD 85628 B, however from its
inferred absolute magnitude $M_G$ $\simeq$ 7.9, it would compare well to 
low-mass solar neighbors like $\eta$ Cas B (K7V: $M_G$ = 7.89) or HD 79210 (M0V;
$M_G$ = 7.96). We predict HD 85628 B to be a $\sim$K8V star with mass $\sim$0.6 \Msun. 

For component masses of 1.75 \Msun\, and 0.6 \Msun, and fiducially setting the observed separation 736 AU as a first estimate of the semi-major axis and assuming $e$ = 0, one would predict an orbital period of $\sim$13000 years and relative orbital motion of 1.68 \kms. This is remarkably similar to the observed difference in tangential motion (1.74\,$\pm$\,0.09 \kms) observed for A and B using the Gaia DR2 astrometry. This calculation also further strengthens the argument that the two stars are likely to be a bound pair.\\

\begin{table*}[htbp]
\centering
\caption{Stellar Parameters}
\begin{tabular}{lcccl}
Parameter  & Unit & HD 85628 A & HD 85628 B & References\\ \hline 
$\alpha$(ICRS)$^1$ & - & 147.58006673040 & 147.57796550484 & GaiaDR2\\
$\delta$(ICRS)$^1$ & - & -66.11392534234 & -66.11477795490 & GaiaDR2\\
Parallax $\varpi$ & mas & 5.8297\,$\pm$\,0.0318 & 5.9508\,$\pm$\,0.0366 & GaiaDR2\\
Spec. Type & - & A3V & ...$^2$ & \citet{Houk75}\\
Dist. & pc &  171.54\,$\pm$\,0.94 & 168.04\,$\pm$\,1.03 & 1/$\varpi$ (GaiaDR2)\\
Proper Motion $\alpha$ $\mu_{\alpha}$ & mas/yr & 6.051\,$\pm$\,0.055 & 5.856\,$\pm$\,0.066 & GaiaDR2\\
Proper Motion $\delta$ $\mu_{\delta}$ & mas/yr & -15.398\,$\pm$\,0.051 & -13.252\,$\pm$\,0.060 & GaiaDR2\\
$G$   & mag & 8.1740 & 14.0490 & GaiaDR2\\
$B_p$ & mag & 8.2832 & 13.6699 & GaiaDR2\\
$R_p$ & mag & 8.0285 & 12.8495 & GaiaDR2\\ 
$B_T$ & mag & 8.442\,$\pm$\,0.009 & ... & \citep{ESA97}\\
$V_T$ & mag & 8.222\,$\pm$\,0.010 & ... & \citep{ESA97}\\
$V$   & mag & 8.19\,$\pm$\,0.01   & ... & \citep{ESA97}\\
$B-V$ & mag & 0.200\,$\pm$\,0.013 & ... & \citep{ESA97}\\
$J$   & mag & 7.837\,$\pm$\,0.021 & ... & \citep{Skrutskie06}\\
$H$   & mag & 7.785\,$\pm$\,0.053 & ... & \citep{Skrutskie06}\\
$K_s$ & mag & 7.750\,$\pm$\,0.023 & ... & \citep{Skrutskie06}\\
$W1$  & mag & 7.646\,$\pm$\,0.024 & ... & \citep{Cutri12}$^3$\\
$W2$  & mag & 7.690\,$\pm$\,0.020 & ... & \citep{Cutri12}$^3$\\
$W3$  & mag & 7.706\,$\pm$\,0.016 & ... & \citep{Cutri12}$^3$\\
$W4$  & mag & 7.619\,$\pm$\,0.074 & ... & \citep{Cutri12}$^3$\\
\hline
\end{tabular}
\caption*{Notes: \\
$^1$ Position epoch is 2015.5 (Gaia DR2). \\
$^2$ Absolute $G$ magnitude consistent with K7-M0 dwarf, but no spectral information available.\\
$^3$ For further information on WISE photometry \cite{Cutri12}, see \citet{Wright10}.}
\label{Tab:stars}
\end{table*}

\section{Discussion and Conclusion}
\label{Sec:Discussion}

%Basic information on the planet here
The parameters describing the MASCARA-4\,b/bRing-1\,b system are shown in Table \ref{Tab:full_param}. We find that MASCARA-4\,b/bRing-1\,b orbits its host star with a period of \periodmasc \space days at a distance of \semiaxismasc \space AU. It has a radius of \radiusmasc \space $R_{\rm{Jup}}$ with a mass of \massmasc \space $M_{\rm{Jup}}$. The planet equilibrium temperature is \tempmasc \space K, assuming a Bond albedo of zero. 

\begin{table*}[htbp]
\centering
\caption{Parameters describing the MASCARA-4\,b/bRing-1\,b system. Due to the reduced chi-square on the photometric fit parameters, the errors on the planet parameters are likely underestimated by $\sim 20\%$.}
\begin{tabular}{lccr}
Parameter & Symbol & Units & Value \\ \hline \hline
Stellar Parameters & & &  \\ \hline
Effective temperature & $T_{\rm{eff}}$ & K & \tempstar \\
Luminosity & $L$ & L$_{\odot}$ & 12.23\,$\pm$\,0.655\\ 
Systemic Velocity & $\gamma$ & km/sec & \sysvelstar \\
Projected rotation speed & $v\sin{i}$ & km/sec & \vsinistar \\
Surface Gravity & log $g$ & - & 4.1\\
Metallicity & [Fe/H] & - & $\sim$0 \\
Stellar Mass & $M_\star$ & $M_\odot$ & \stellarmass \\
Stellar Radius  & $R_\star$ & $R_\odot$ & \stellarradius \\
Age & $\tau$ & Gyr & $\sim$0.8\\ \hline
Planet Parameters &  &  &  \\ \hline
Planet radius & $R_P$ & R$_{\rm{Jup}}$ & \radiusmasc\\
Planet mass & $M_P$ & M$_{\rm{Jup}}$ &  \massmasc\\
Equilibrium temperature$^a$ & $T_{\rm{eq}}$ &  K & \tempmasc  \\ \hline
System Parameters &  &  &  \\ \hline
Epoch & $T_P$ & BJD & \epochmasc \\
Period & $P$ & days & \periodmasc  \\
Semi-major axis &  $a$  & AU & \semiaxismasc \\
Eccentricity & $e$ & - & 0 (fixed) \\
Projected Obliquity & $|\lambda|$ & \textdegree  &  \obliquitymasc\\ \hline
\end{tabular}
\caption*{ Notes: $^a$ Assuming a Bond albedo of 0.}
\label{Tab:full_param}
\end{table*}

%About the system/planet

The planet is in a retrograde orbit, $|\lambda|=$\obliquitymasc \textdegree, around its host, a hot (T$_{\rm{eff}}=$\tempstar \space K), early-type (A3V) star. It is the brightest system known to date with a planet in a retrograde orbit ($112.5$\textdegree$\leq\  |\lambda|\leq247.5$ \textdegree, following \cite{Addison2013ApJ...774L...9A}), and currently one of only fifteen known retrograde systems \citep{tepcet2011MNRAS.417.2166S}. This is contrasted with 18 exoplanets in known polar orbits ($67.5$\textdegree$ < |\lambda|<112.5$ \textdegree), with 219 currently known exoplanets with derived obliquities. As found in earlier work \citep{Albrecht2012,Schlaufman2010ApJ...719..602S,Winn2010ApJ...718L.145W}, hot Jupiters orbiting early type stars have a significantly higher probability to be in a polar or retrograde orbit. Currently, 0.081\% of planets orbiting stars with $T_{\rm{eff}} < 6300 K$ are found in such orbits, while this is 39.6\% for planets those orbiting stars with $T_{\rm{eff}} > 6300 K$. It is unclear how such misalignment could result from early planet-disk interactions, and it is more likely that this is caused by Kozai-Lidov oscillations with a more distant companion. Indeed, we find HD\,85628 to have a late-type companion (likely a $\sim$K8 dwarf), currently at a projected distance of $\sim$736 AU. It is yet unclear why hot Jupiters orbiting early type stars are more likely to be misaligned. It is possible that their formation mechanism is different from current models. However, it has also been suggested that the fact that outer layers of lower mass stars are convective, which shortens the time scale of orbital re-alignment \citep{Winn2010ApJ...718L.145W}, erasing the evidence of the Kozai-Lidov effect.   

The Doppler shadow measurement only determined the projected spin-orbit angle, while the real spin-orbit angle is degenerate with the stellar inclination. The \vsini\, of the host star is relatively low (\vsinistar\,\kms) for its spectral type, implying a low inclination, making its true de-projected orbit either more retrograde, or more polar $-$ depending on the sign of the inclination angle.

\begin{figure*}
    \centering
    \includegraphics[scale=0.4]{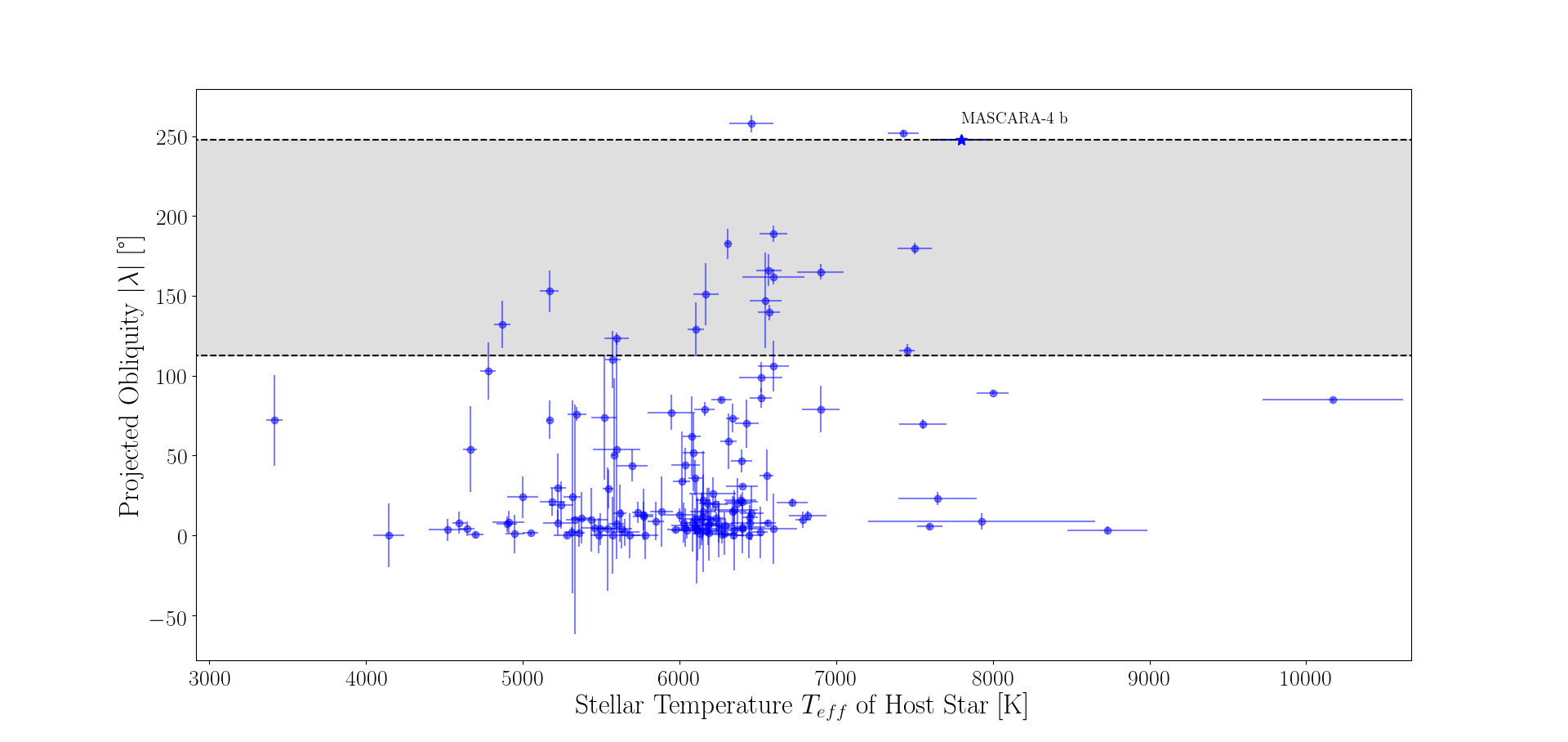}
    \caption{All currently known exoplanets with measured projected obliquities. as taken from, the TEPCat catalogue \citep{tepcet2011MNRAS.417.2166S}. MASCARA-4\,b is shown as a blue star. It further emphasizes the trend that those planets orbiting hot stars have a larger probability to be in a mis-aligned orbit. The grey shaded regions denote what is considered to be a retrograde planet, with projected obliquity $112.5$\textdegree$\leq |\lambda| \leq 247.5$\textdegree. }
    \label{fig:planet_obliquities}
\end{figure*}

The brightness of the host star, combined with the short orbital period and high planet temperature, make MASCARA-4\,b an excellent candidate for follow-up atmospheric studies, reminiscent of WASP-33\,b \citep{WASP33b_2010MNRAS.407..507C}. The hottest gas giant planets are found to have unique atmospheric features. The hottest of all, KELT-9\,b \citep{KELT9b_2017Natur.546..514G,YanHenningKelt92018NatAs...2..714Y}, exhibits an optical transmission spectrum rich of metallic atoms and ions such as Iron and Titanium \citep{KELT9-IRON_TITANIUM2018Natur.560..453H}, while it shows strong H$\alpha$ absorption originating from a halo of escaping hydrogen gas. A planet like WASP-33\,b shows a thermal dayside spectrum dominated by TiO emission lines, indicative of a strong thermal inversion \citep{Nugroho2017AJ....154..221N}. We will require detailed atmospheric observations of a significant sample of the hottest Jupiters to understand how these features depend on the planet and possible host star properties, and to understand the underlying physical and chemical mechanisms. In any case, these systems will be the most easy to study with current instruments, and with future observatories such as the James Webb Space Telescope and the ground-based extremely large telescopes.   

The NASA TESS satellite is covering this target in Sector 9, which may constrain the secondary eclipse and phase curve of the planet, and thus its dayside temperature and global circulation. 
\begin{acknowledgements}

I.S. acknowledges support from a NWO VICI grant (639.043.107).
This project has received funding from the European Research Council (ERC) under the European Union’s Horizon 2020 research and innovation programme (grant agreement nr. 694513). 
E.E.M. and S.N.M. acknowledge support from the NASA NExSS programme. 
SNM is a U.S. Department of Defense SMART scholar sponsored by the U.S. Navy through SSC-LANT.
EEM acknowledges support from the NASA NExSS program and a JPL RT\&D award. 
A.W acknowledges the support of the SNSF by the grant number P2GEP2 178191
Part of this research was carried out at the Jet Propulsion Laboratory, California Institute of Technology, under a contract with the National Aeronautics and Space Administration. 
This work has made use of data from the European Space Agency (ESA) mission Gaia (https://www.cosmos.esa.int/gaia), processed by the Gaia Data Processing and Analysis Consortium (DPAC,https://www.cosmos.esa.int/web/ gaia/dpac/consortium).
This research has made use of the SIMBAD database, operated at CDS, Strasbourg, France. 
This research has made use of the VizieR catalogue access tool, CDS, Strasbourg, France. 
We have benefited greatly from the publicly available programming language Python, including the numpy \citep{numpy}, matplotlib \citep{MatplotlibHunter:2007}, pyfits, scipy \citep{Scipy} and h5py \citep{H5py} packages. Pyfits is a product of the Space Telescope Science Institute, which is operated by AURA for NASA.
The authors would like to acknowledge the support staff at both the South African Astronomical Observatory and Siding Spring Observatory for keeping both bRing stations maintained and running, and the ESO La Silla TMES staff for their support with MASCARA-S.
\end{acknowledgements}

%-------------------------------------------------------------------

\bibliography{MASCARA_4}
\bibliographystyle{aa}

\end{document}